\documentclass[prd,onecolumn]{revtex4}
\usepackage{dcolumn}
\usepackage{multirow}
\usepackage{graphicx}
\usepackage{amssymb}
\usepackage{bm}
\usepackage{hyperref}%for .pdf
\usepackage{epstopdf}%for .pdf
\usepackage{color}
\usepackage{mathrsfs}
\usepackage{amsmath,amssymb,amsthm}
\usepackage{rotating}
\usepackage{sverb, longtable}
\usepackage{subfigure}

\usepackage[graphicx]{realboxes}
\usepackage{adjustbox}
\begin{document}

\title{Shaving the Hair of Black Hole with Sagittarius A$^*$ from Event Horizon Telescope}
%Testing the No-Hair Theorem with Sagittarius A$^*$ from Event Horizon Telescope
\author{Deng Wang}
\email{cstar@nao.cas.cn}
\affiliation{National Astronomical Observatories, Chinese Academy of Sciences, Beijing, 100012, China}
\begin{abstract}
Recently, the Event Horizon Telescope collaboration has reported the first image of the supermassive black hole Sagittarius in the Galactic Center. We attempt to test the validity of the no-hair theorem of black holes using this new shadow observation.  
Considering the Einstein-Maxwell-klein-Gordon theory with a minimally-coupled scalar field, we find that our numerical result is consistent with the prediction of the no-hair theorem. However, we can not rule out the possibility that black holes with scalar hair may exist in some special cases.

\end{abstract}
\maketitle

\section{Introduction}
As the most important compact objects predicted by general relativity \cite{Einstein1915}, black holes, play a basic role in fundamental physics. Since firstly studied by Schwarzschild, black holes always lie in the cores of various theoretical researches when one attempts to unify quantum mechanics with general relativity. The typical property of a black hole is that it has an event horizon, within which a light ray can not escape from the tight binding of extremely strong gravity \cite{Schwarzschild1916}. In astrophysical observations, although black holes have been found in a wide range of masses including stellar-mass ones \cite{Webster1972,Remillard:2006fc} and supermassive ones \cite{Lynden-Bell:1969gsv,Kormendy:1995er,Miyoshi:1995da}, we lack the phenomenological knowledge of their event horizons in observations over a long period of time. Until recent several years, two new observational approaches permit us to explore the physical processes and spacetime geometry around the event horizon of a black hole. The former is more and more mature gravitational wave observations from the LIGO and VIRGO collaboration \cite{LIGOScientific:2016aoc}, while the latter is the imaging of black hole shadow from the Event Horizon Telescope (EHT) \cite{EventHorizonTelescope:2019dse},  which is a global very long baseline interferometry working at a wavelength of 1.3 mm (230 GHz). Especially, the EHT imaging survey provides an excellent opportunity for us to study the dark shadow induced by gravitational light bending and photon capture at the event horizon.             

In 2019, it is very exciting that the EHT collaboration firstly reported the shadow of the supermassive black hole M$87^\star$ residing in the giant elliptical galaxy Messier 87 \cite{EventHorizonTelescope:2019dse}. In light of this ground-breaking progress, various fundamental physics have been extensively investigated \cite{Moffat:2019uxp,Giddings:2019jwy,Wei:2019pjf,Davoudiasl:2019nlo,Bambi:2019tjh,Vagnozzi:2019apd,Long:2019nox,Contreras:2019cmf,Neves:2019lio,Tian:2019yhn,Banerjee:2019nnj,Shaikh:2019hbm,Kumar:2019pjp,Allahyari:2019jqz,Yan:2019hxx,Jusufi:2019ltj,Rummel:2019ads,Li:2020drn,Konoplya:2020bxa,Guo:2020zmf,Wei:2020ght,Creci:2020mfg,Chen:2019fsq, Shaikh:2021yux,Hu:2020usx,Chowdhuri:2020ipb,Afrin:2021imp}. In particular, the no-hair theorem (NHT) of black holes, which states that all black hole spacetime configurations of the Einstein-Maxwell equations of gravitation and electromagnetism in general relativity can be completely described by only three classical parameters: mass, angular momentum, and electric charge, can be effectively tested by using the shadow size of M$87^\star$ \cite{Khodadi:2020jij}. At present, it is interesting to study the validity of NHT based on three reasons: (i) the community wonders urgently if black holes have other external information except for above three parameters; (ii) the coming of EHT imaging observations has improved to a large extent our understanding of near-horizon properties of black holes; (iii) the AdS/CFT correspondence \cite{Maldacena:1997re} requires a deep study on the behaviors of matter fields near the charged black holes \cite{Gubser:2005ih}. In this theoretical direction, Ref.\cite{Khodadi:2020jij} has explored the possible violations of NHT for black holes with scalar hair by using the shadow of M$87^\star$. Most recently, the EHT collaboration \cite{Akiyama2022} has released the latest image of Sagittarius A$^\star$ in the center of our Milk Way galaxy, and we are motivated by testing the validity of NHT by using this new observation. We find that our result agrees well with the standard NHT statement, but it is still possible to violate the NHT in some special cases.  

This work is outlined in the following manner. In the next section, we review briefly the theory, black hole solutions and their shadows. In Section III, we describe the data from EHT and show the analysis results. Discussions and conclusions are presented in the final section. The units $c=G=\hbar=1$ are taken in this work.

\section{Theory}
In this study, we consider the Einstein-Maxwell-klein-Gordon theory with a minimally-coupled scalar field $\phi$ and its action reads as \cite{Martinez:2006an,Anabalon:2009qt,Xu:2013nia,Anabalon:2013qua,Fan:2015oca,Gonzalez:2013aca,Gonzalez:2014tga,Hui:2019aqm}
\begin{equation}
S = \int d^4x\sqrt{-g}\left[\frac{R}{2\kappa}-\frac{1}{4}F_{\mu\nu}F^{\mu\nu}-\frac{1}{2}g^{\mu\nu}\Delta_\mu\phi \Delta_\nu\phi-V(\phi)\right], \label{1}
\end{equation}
where $\kappa\equiv8\pi$, $R$ is Ricci scalar, $g_{\mu\nu}$ the metric tensor, $V(\phi)$ the potential, and the electromagnetic field tensor $F_{\mu_\nu}=\partial_\mu A_\nu-\partial_\nu A_\mu$, where the four-potential $A_\mu\equiv(A_t(r),0,0,0)$. Varying the action with respect to the metric, one can express the extended gravitational field equation as  
\begin{equation}
R_{\mu\nu}-\frac{1}{2}g_{\mu\nu}R = \kappa\left[T_{\mu\nu}^{(F)}+T_{\mu\nu}^{(\phi)}\right],  \label{2}
\end{equation}
where $T_{\mu\nu}^{(F)}$ and $T_{\mu\nu}^{(\phi)}$ are the energy-momentum tensors of electromagnetic and scalar fields, respectively. Subsequently, minimizing the action with respect to the scalar field and four-potential, one can obtain, respectively, the so-called Klein-Gordon equation and Maxwell equation as 
\begin{equation}
\Box\phi=\frac{dV}{d\phi}, \label{3}
\end{equation}
\begin{equation}
\Delta\nu F^{\mu\nu}=0.  \label{4}
\end{equation}
In order to study the shadow of a black hole in this theory, we will consider a static spherical symmetric solution. In Ref.\cite{Khodadi:2020jij}, this approximation has been verified to be valid based on the present EHT imaging precision, when applying such a solution to a realistic astrophysical black hole with rotation. We shall take the following metric
\begin{equation}
ds^2=-f(r)dt^2+\frac{1}{f(r)}dr^2+c^2(r)d\Omega^2, \label{5}
\end{equation}       
where $f(r)$ is the metric function and the origin factor $r^2$ in general relativity is replaced by $c^2(r)$ in the extended theory. To obtain the final expression of $f(r)$, we should insert four functions $\phi(r)$, $c(r)$, $A_t(r)$ and $V(\phi)$ by hand according to some physically well-behaved properties. Specifically, we choose them as
\begin{equation}
\phi(r)=\frac{1}{\sqrt{2}}\mathrm{ln}\left(\frac{\nu}{r}+1\right), \label{6}
\end{equation}
\begin{equation}
c(r)=\sqrt{r(r+\nu)}, \label{7}
\end{equation}
\begin{equation}
A_t(r)=\frac{Q}{\nu}\mathrm{ln}\left(\frac{r}{r+\nu}\right), \label{8}
\end{equation}
\begin{equation}
V(\phi)=\frac{g(\phi)}{2\nu^4}e^{-2\sqrt{2}\phi}, \label{9}
\end{equation}
where $\nu$ denotes the hair parameter, $Q$ is the electronic charge of a black hole, and $g(\phi)$ is given by Eq.(2.16) in Ref.\cite{Khodadi:2020jij}. By substituting Eq.(\ref{5}) into Eq.(\ref{2}) and using Eqs.(\ref{3}-\ref{8}), one can get $f(r)$ for the minimally-coupled charged black holes:
\begin{equation}
f(r) = -\frac{6M(\nu+2r)+2\nu r+2Q^2}{\nu^2}-
\mathrm{ln}\left(\frac{r}{\nu+r}\right)
\left[\frac{2Q^2(\nu+2r)+2r(\nu+r)(\nu+6M)}{\nu^3}+
\frac{2rQ^2(r+\nu)\mathrm{ln}\left(\frac{r}{\nu+r}\right)}{\nu^4}\right]. \label{10}
\end{equation}  
To calculate the shadow cast by the above black hole solution, we shall consider the photon geodesics parameterized by $x^\mu(\tau)$ ($\tau$ is the affine parameter), and the Lagrangian of photon geodesics reads as  
\begin{equation}
\mathcal{L}=-f(r)\dot{t}^2+\frac{1}{f(r)}\dot{r}^2+c^2(\dot{\theta}^2+\dot{\phi}^2), \label{11}
\end{equation}
where the dot is the derivative with respect to $\tau$. Furthermore, the equations of motion in the equatorial plane ($\theta=\pi/2$) are shown as 
\begin{equation}
E=f(r)\dot{t}, \label{12}
\end{equation}
\begin{equation}
L=c^2\dot{\phi}, \label{13}
\end{equation}
\begin{equation}
f(r)\dot{t}^2-\frac{1}{f(r)}\left(\frac{dr}{d\phi}\right)^2\dot{\phi}^2-c^2\dot{\phi}^2=0, \label{14}
\end{equation}
where $E$ and $L$ denote the energy and angular momentum of photons. Substituting Eqs.(\ref{12}) and (\ref{13}) into Eq.(\ref{14}), one can have the equation of radial motion of photons
\begin{equation}
F(r)\equiv\left(\frac{dr}{d\phi}\right)^2=c^4(r)\left[\frac{E^2}{L^2}-\frac{f(r)}{c^2(r)}\right], \label{15}
\end{equation}
By studying the photon sphere of the considered black hole and implementing the standard analysis procedure in computing the shadow size of a black hole, one can obtain the cosine value of the angle $\psi$ between the four-vector $k^\mu$ tangent to the photon's path and the position vector $D^\mu$ of a distant observer located at $r=r_0$,  
\begin{equation}
\mathrm{cos\,\psi}=\left[\frac{F(r)}{F(r)+c^2(r)f(r)}\right]^{\frac{1}{2}}. \label{16}
\end{equation}
The angle $\psi$ is actually a function of the parameters $Q$, $\nu$ and radial coordinate $r$, i.e., $\psi=\psi(r,Q,\nu)$. Furthermore, the shadow radius can be easily expressed as $r_{sh}=r_0\mathrm{tan}\,\psi$. One can compare this theoretical prediction with the EHT observations.

\begin{figure}
	\centering
	\includegraphics[scale=0.6]{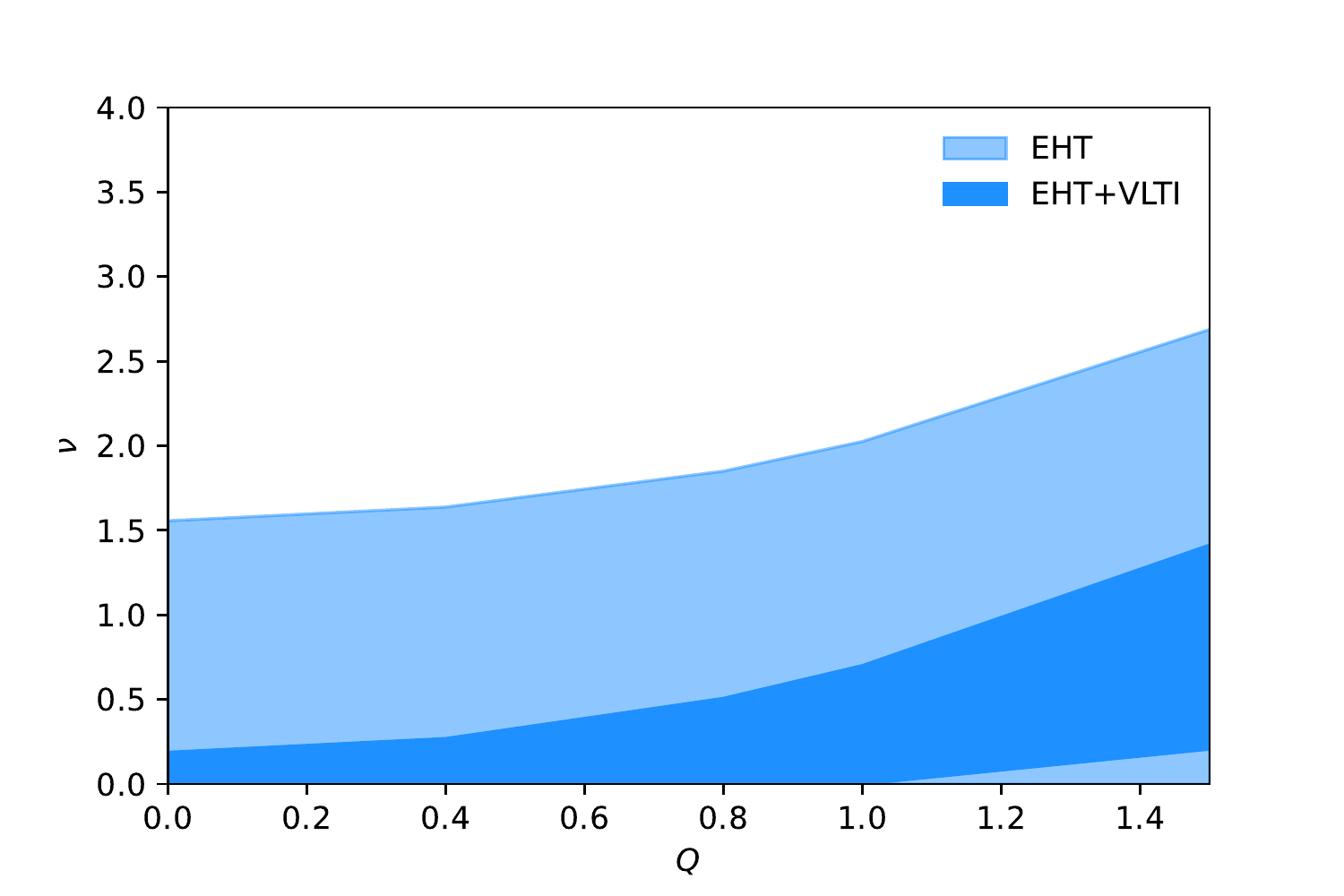}
	\caption{The 2-dimensional $(Q,\nu)$ parameter space constrained by the EHT imaging observations. Here we assume $\nu\geq0$.}\label{f1}
\end{figure}

\section{Data and results} 
In 2019, the EHT collaboration resolved the compact radio source M87$^\star$ as an asymmetric bright emission ring with a diameter of $42\pm3$ $\mu$as, and did not give the estimated angular shadow diameter. Most recently, for Sagittarius A$^\star$, they not only release the bright, thick emission ring with a diameter of $51.8\pm2.3$ $\mu$as but also give the estimated angular shadow diameter $48.7\pm7.0$ $\mu$as (see Table.1 in Ref.\cite{Akiyama2022}). It is easy to find that the shadow size has a larger uncertainty than the ring size. As a consequence, we will conservatively adopt the angular shadow diameter $\theta=48.7\pm7.0$ $\mu$as to test the NHT for Sagittarius A$^\star$ in this study. This can ensure the validity of our analysis. 

Combining the data from EHT with the distance $8150\pm150$ pc measured from VLBI \cite{Akiyama2022}, we obtain the angular diameter $9.92\pm3.01$ at $1\,\sigma$ level for Sagittarius A$^\star$ by redefining the measured angular diameter as $d_{Sgr A^\star}\equiv D\theta/M$, where $D$ denotes the distance from Sagittarius A$^\star$ to us. Note that we have shown the angular size in units of mass. One can easily find that, in light of current EHT precision, the data is well compatible with the Schwarzschild solution, whose $d=6\sqrt{3}$. However, the error is large, since the black hole mass $4.0^{+1.1}_{-0.6}\times10^{6}$ M$_\odot$ has a large uncertainty. To improve the constraining power, we combine the VLTI mass and distance \cite{GRAVITY:2021xju} (see Table.1 in Ref.\cite{Akiyama2022}) with EHT shadow size (hereafter EHT+VLTI) to implement the analysis. We obtain $9.38\pm1.34$ at $1\,\sigma$ level and, as expected, the uncertainty is improved from $30.3\%$ to $14.3\%$.    

As stated in the above section, to test the NHT, we compare the analysis result with theory by mapping the constrained shadow angular diameter with errors into the 2-dimensional space $(Q,\nu)$, and the corresponding result is presented in Fig.\ref{f1}. We find that the EHT-only data can just give upper bounds on the scalar hair parameter $\nu$ over any electronic charge, but if combined with VLTI, the constraining power will be obviously enhanced. We observe that EHT+VLTI can reduce the original $(Q,\nu)$ parameter space from EHT by at least $50\%$, and that the constraining power of EHT and EHT+VLTI both decrease with increasing $Q$. The growth of constraining power between EHT+VLTI and EHT is the strongest when $Q=0$, namely, EHT+VLTI gives $\nu<0.19$. Interestingly, EHT+VLTI starts to give the lower bound on $\nu$ when $Q\geq1$. For example, $\nu>0.21$ when $Q=1.5$. 

It is clear that our result is consistent with NHT at $1\,\sigma$ level, since the constrained parameter space permits the existence of $\nu=0$. However, it is interesting that the violation of NHT is possible based on the fact that the lower bound on $\nu$ will be larger than 0 when $Q\geq1$.

\section{Discussions and conclusions}
Recently, the imaging survey of the Event Horizon Telescope collaboration has inspired the community to exploring the black hole physics by using the near-horizon scale observations. In light of the latest image data of Sagittarius A$^\star$, we attempt to investigate the validity of NHT. 

Using the EHT-only data, we find that the constraint on the 2-dimensional parameter space $(Q,\nu)$ is relatively loose. However, if combing EHT with VLTI data, the constraining power can be improved by at least a factor of 2 over the whole $Q$ range. This implies that currently observational precision of EHT is limited. The best result we obtain is the $1\,\sigma$ bound $\nu<0.19$ when there is no electronic charge. 

Overall, our constraining result is consistent with NHT, since the constrained parameter space allow the existence of $\nu=0$. Nonetheless, very interestingly, we do not rule out the possibility that NHT may be violated, because the lower bound of the hair parameter $\nu$ starts to be larger than 0 when $Q\geq1$. 

It is worth noting that our constraint on NHT is affected to a large extent by the value of electronic charge. In order to obtain more information about the test of NHT, one should perform the constraint by considering a 2-dimensional parameter space $(Q,\nu)$ instead of the traditional one-parameter method. We believe that future EHT imaging data with high precision will help test the NHT and probing the fundamental physics better. 

\section{Acknowledgments}
DW thank Changjun Gao and Yuan Sun for insightful discussions during the preparation of the manuscript. This work is supported by National Natural Science Foundation of China under Grants No.11988101 and No.11851301.


\begin{thebibliography}{99}
\bibitem{Einstein1915}
A. Einstein, Sitzungsberichte der Königlich Preußischen Akademie der
Wissenschaften (Berlin: Deutsche Akademie der Wissenschaften zu Berlin), (1915).

\bibitem{Schwarzschild1916}
K. Schwarzschild, AbhKP, 189, (1916).

\bibitem{Webster1972}
B. Webster and P. Murdin, Nature, {\bf 235}, 37 (1972).

%\cite{Remillard:2006fc}
\bibitem{Remillard:2006fc}
R.~A.~Remillard and J.~E.~McClintock,
%``X-ray Properties of Black-Hole Binaries,''
Ann. Rev. Astron. Astrophys. \textbf{44}, 49-92 (2006).

%\cite{Lynden-Bell:1969gsv}
\bibitem{Lynden-Bell:1969gsv}
D.~Lynden-Bell,
%``Galactic nuclei as collapsed old quasars,''
Nature \textbf{223}, 690 (1969).

%\cite{Kormendy:1995er}
\bibitem{Kormendy:1995er}
J.~Kormendy and D.~Richstone,
%``Inward bound: The Search for supermassive black holes in galactic nuclei,''
Ann. Rev. Astron. Astrophys. \textbf{33}, 581 (1995).

%\cite{Miyoshi:1995da}
\bibitem{Miyoshi:1995da}
M.~Miyoshi {\it et al.},
%``Evidence for a black hole from high rotation velocities in a subparsec region of NGC4258,''
Nature \textbf{373}, 127-129 (1995).

%\cite{LIGOScientific:2016aoc}
\bibitem{LIGOScientific:2016aoc}
B.~P.~Abbott \textit{et al.} [LIGO Scientific and Virgo],
%``Observation of Gravitational Waves from a Binary Black Hole Merger,''
Phys. Rev. Lett. \textbf{116}, no.6, 061102 (2016).

%\cite{EventHorizonTelescope:2019dse}
\bibitem{EventHorizonTelescope:2019dse}
K.~Akiyama \textit{et al.} [Event Horizon Telescope],
%``First M87 Event Horizon Telescope Results. I. The Shadow of the Supermassive Black Hole,''
Astrophys. J. Lett. \textbf{875}, L1 (2019).



%\cite{Moffat:2019uxp}
\bibitem{Moffat:2019uxp}
J.~W.~Moffat and V.~T.~Toth,
%``Masses and shadows of the black holes Sagittarius A* and M87* in modified gravity,''
Phys. Rev. D \textbf{101}, no.2, 024014 (2020).

%\cite{Giddings:2019jwy}
\bibitem{Giddings:2019jwy}
S.~B.~Giddings,
%``Searching for quantum black hole structure with the Event Horizon Telescope,''
Universe \textbf{5}, no.9, 201 (2019).


%\cite{Wei:2019pjf}
\bibitem{Wei:2019pjf}
S.~W.~Wei, Y.~C.~Zou, Y.~X.~Liu and R.~B.~Mann,
%``Curvature radius and Kerr black hole shadow,''
JCAP \textbf{08}, 030 (2019).

%\cite{Davoudiasl:2019nlo}
\bibitem{Davoudiasl:2019nlo}
H.~Davoudiasl and P.~B.~Denton,
%``Ultralight Boson Dark Matter and Event Horizon Telescope Observations of M87*,''
Phys. Rev. Lett. \textbf{123}, no.2, 021102 (2019).

%\cite{Bambi:2019tjh}
\bibitem{Bambi:2019tjh}
C.~Bambi, K.~Freese, S.~Vagnozzi and L.~Visinelli,
%``Testing the rotational nature of the supermassive object M87* from the circularity and size of its first image,''
Phys. Rev. D \textbf{100}, no.4, 044057 (2019).

%\cite{Vagnozzi:2019apd}
\bibitem{Vagnozzi:2019apd}
S.~Vagnozzi and L.~Visinelli,
%``Hunting for extra dimensions in the shadow of M87*,''
Phys. Rev. D \textbf{100}, no.2, 024020 (2019).

%\cite{Long:2019nox}
\bibitem{Long:2019nox}
F.~Long, J.~Wang, S.~Chen and J.~Jing,
%``Shadow of a rotating squashed Kaluza-Klein black hole,''
JHEP \textbf{10}, 269 (2019).

%\cite{Contreras:2019cmf}
\bibitem{Contreras:2019cmf}
E.~Contreras, \'A.~Rinc\'on, G.~Panotopoulos, P.~Bargue\~no and B.~Koch,
%``Black hole shadow of a rotating scale--dependent black hole,''
Phys. Rev. D \textbf{101}, no.6, 064053 (2020).

%\cite{Neves:2019lio}
\bibitem{Neves:2019lio}
J.~C.~S.~Neves,
%``Upper bound on the GUP parameter using the black hole shadow,''
Eur. Phys. J. C \textbf{80}, no.4, 343 (2020).

%\cite{Tian:2019yhn}
\bibitem{Tian:2019yhn}
S.~X.~Tian and Z.~H.~Zhu,
%``Testing the Schwarzschild metric in a strong field region with the Event Horizon Telescope,''
Phys. Rev. D \textbf{100}, no.6, 064011 (2019).

%\cite{Banerjee:2019nnj}
\bibitem{Banerjee:2019nnj}
I.~Banerjee, S.~Chakraborty and S.~SenGupta,
%``Silhouette of M87*: A New Window to Peek into the World of Hidden Dimensions,''
Phys. Rev. D \textbf{101}, no.4, 041301 (2020).

%\cite{Shaikh:2019hbm}
\bibitem{Shaikh:2019hbm}
R.~Shaikh and P.~S.~Joshi,
%``Can we distinguish black holes from naked singularities by the images of their accretion disks?,''
JCAP \textbf{10}, 064 (2019).

%\cite{Kumar:2019pjp}
\bibitem{Kumar:2019pjp}
R.~Kumar, S.~G.~Ghosh and A.~Wang,
%``Shadow cast and deflection of light by charged rotating regular black holes,''
Phys. Rev. D \textbf{100}, no.12, 124024 (2019).

%\cite{Allahyari:2019jqz}
\bibitem{Allahyari:2019jqz}
A.~Allahyari, M.~Khodadi, S.~Vagnozzi and D.~F.~Mota,
%``Magnetically charged black holes from non-linear electrodynamics and the Event Horizon Telescope,''
JCAP \textbf{02}, 003 (2020).

%\cite{Yan:2019hxx}
\bibitem{Yan:2019hxx}
S.~F.~Yan {\it et al.},
%``Testing the equivalence principle via the shadow of black holes,''
Phys. Rev. Res. \textbf{2}, no.2, 023164 (2020).

%\cite{Jusufi:2019ltj}
\bibitem{Jusufi:2019ltj}
K.~Jusufi,
%``Quasinormal Modes of Black Holes Surrounded by Dark Matter and Their Connection with the Shadow Radius,''
Phys. Rev. D \textbf{101}, no.8, 084055 (2020).

%\cite{Rummel:2019ads}
\bibitem{Rummel:2019ads}
M.~Rummel and C.~P.~Burgess,
%``Constraining Fundamental Physics with the Event Horizon Telescope,''
JCAP \textbf{05}, 051 (2020).

%\cite{Li:2020drn}
\bibitem{Li:2020drn}
P.~C.~Li, M.~Guo and B.~Chen,
%``Shadow of a Spinning Black Hole in an Expanding Universe,''
Phys. Rev. D \textbf{101}, no.8, 084041 (2020).

%\cite{Konoplya:2020bxa}
\bibitem{Konoplya:2020bxa}
R.~A.~Konoplya and A.~F.~Zinhailo,
%``Quasinormal modes, stability and shadows of a black hole in the 4D Einstein\textendash{}Gauss\textendash{}Bonnet gravity,''
Eur. Phys. J. C \textbf{80}, no.11, 1049 (2020).


%\cite{Guo:2020zmf}
\bibitem{Guo:2020zmf}
M.~Guo and P.~C.~Li,
%``Innermost stable circular orbit and shadow of the $4D$ Einstein\textendash{}Gauss\textendash{}Bonnet black hole,''
Eur. Phys. J. C \textbf{80}, no.6, 588 (2020).

%\cite{Wei:2020ght}
\bibitem{Wei:2020ght}
S.~W.~Wei and Y.~X.~Liu,
%``Testing the nature of Gauss-Bonnet gravity by four-dimensional rotating black hole shadow,''
Eur. Phys. J. Plus \textbf{136}, no.4, 436 (2021).

%\cite{Chen:2019fsq}
\bibitem{Chen:2019fsq}
Y.~Chen, J.~Shu, X.~Xue, Q.~Yuan and Y.~Zhao,
%``Probing Axions with Event Horizon Telescope Polarimetric Measurements,''
Phys. Rev. Lett. \textbf{124}, no.6, 061102 (2020)

%\cite{Creci:2020mfg}
\bibitem{Creci:2020mfg}
G.~Creci, S.~Vandoren and H.~Witek,
%``Evolution of black hole shadows from superradiance,''
Phys. Rev. D \textbf{101}, no.12, 124051 (2020).



%\cite{Shaikh:2021yux}
\bibitem{Shaikh:2021yux}
R.~Shaikh, K.~Pal, K.~Pal and T.~Sarkar,
%``Constraining alternatives to the Kerr black hole,''
Mon. Not. Roy. Astron. Soc. \textbf{506}, no.1, 1229-1236 (2021).

%\cite{Hu:2020usx}
\bibitem{Hu:2020usx}
Z.~Hu, Z.~Zhong, P.~C.~Li, M.~Guo and B.~Chen,
%``QED effect on a black hole shadow,''
Phys. Rev. D \textbf{103}, no.4, 044057 (2021).

%\cite{Chowdhuri:2020ipb}
\bibitem{Chowdhuri:2020ipb}
A.~Chowdhuri and A.~Bhattacharyya,
%``Shadow analysis for rotating black holes in the presence of plasma for an expanding universe,''
Phys. Rev. D \textbf{104}, no.6, 064039 (2021).

%\cite{Afrin:2021imp}
\bibitem{Afrin:2021imp}
M.~Afrin, R.~Kumar and S.~G.~Ghosh,
%``Parameter estimation of hairy Kerr black holes from its shadow and constraints from M87*,''
Mon. Not. Roy. Astron. Soc. \textbf{504}, 5927-5940 (2021).

%\cite{Khodadi:2020jij}
\bibitem{Khodadi:2020jij}
M.~Khodadi, A.~Allahyari, S.~Vagnozzi and D.~F.~Mota,
%``Black holes with scalar hair in light of the Event Horizon Telescope,''
JCAP \textbf{09}, 026 (2020).

%\cite{Maldacena:1997re}
\bibitem{Maldacena:1997re}
J.~M.~Maldacena,
%``The Large N limit of superconformal field theories and supergravity,''
Int.J.Theor.Phys. {\bf38}, 1113 (1999), Adv. Theor. Math. Phys. \textbf{2}, 231-252 (1998).

%\cite{Gubser:2005ih}
\bibitem{Gubser:2005ih}
S.~S.~Gubser,
%``Phase transitions near black hole horizons,''
Class. Quant. Grav. \textbf{22}, 5121-5144 (2005).

\bibitem{Akiyama2022}
K. Akiyama {\it et al.} [Event Horizon Telescope], Astrophys. J. Lett. 930, L12 (2022).



%\cite{Martinez:2006an}
\bibitem{Martinez:2006an}
C.~Martinez and R.~Troncoso,
%``Electrically charged black hole with scalar hair,''
Phys. Rev. D \textbf{74}, 064007 (2006).

%\cite{Anabalon:2009qt}
\bibitem{Anabalon:2009qt}
A.~Anabalon and H.~Maeda,
%``New Charged Black Holes with Conformal Scalar Hair,''
Phys. Rev. D \textbf{81}, 041501 (2010).

%\cite{Xu:2013nia}
\bibitem{Xu:2013nia}
W.~Xu and L.~Zhao,
%``Charged black hole with a scalar hair in (2+1) dimensions,''
Phys. Rev. D \textbf{87}, no.12, 124008 (2013).

%\cite{Anabalon:2013qua}
\bibitem{Anabalon:2013qua}
A.~Anabalon, D.~Astefanesei and R.~Mann,
%``Exact asymptotically flat charged hairy black holes with a dilaton potential,''
JHEP \textbf{10}, 184 (2013).

%\cite{Fan:2015oca}
\bibitem{Fan:2015oca}
Z.~Y.~Fan and H.~Lu,
%``Charged Black Holes with Scalar Hair,''
JHEP \textbf{09}, 060 (2015).

%\cite{Gonzalez:2013aca}
\bibitem{Gonzalez:2013aca}
P.~A.~Gonz\'alez, E.~Papantonopoulos, J.~Saavedra and Y.~V\'asquez,
%``Four-Dimensional Asymptotically AdS Black Holes with Scalar Hair,''
JHEP \textbf{12}, 021 (2013).

%\cite{Gonzalez:2014tga}
\bibitem{Gonzalez:2014tga}
P.~A.~Gonz\'alez, E.~Papantonopoulos, J.~Saavedra and Y.~V\'asquez,
%``Extremal Hairy Black Holes,''
JHEP \textbf{11}, 011 (2014).

%\cite{Hui:2019aqm}
\bibitem{Hui:2019aqm}
L.~Hui, D.~Kabat, X.~Li, L.~Santoni and S.~S.~C.~Wong,
%``Black Hole Hair from Scalar Dark Matter,''
JCAP \textbf{06}, 038 (2019).

%\cite{GRAVITY:2021xju}
\bibitem{GRAVITY:2021xju}
R.~Abuter \textit{et al.} [GRAVITY Collaboration],
%``Mass distribution in the Galactic Center based on interferometric astrometry of multiple stellar orbits,''
Astron. Astrophys. \textbf{657}, L12 (2022).























\end{thebibliography}
\end{document}